\documentclass{iopart}
\usepackage{graphicx}

\usepackage{latexsym}
\usepackage{amssymb}

\begin{document}

\title[]
{Cosmological Perturbations without Inflation}

\author{Fulvio Melia\footnote{John Woodruff Simpson Fellow.}}
\address{Department of Physics, The Applied Math Program, and Department of Astronomy,\\  
The University of Arizona, AZ 85721, USA}

\begin{abstract}
A particularly attractive feature of inflation is that quantum fluctuations in
the inflaton field may have seeded inhomogeneities in the cosmic microwave
background (CMB) and the formation of large-scale structure. In this paper,
we demonstrate that a scalar field with zero active mass, i.e., with an
equation of state $\rho+3p=0$, where $\rho$ and $p$ are its energy density
and pressure, respectively, could also have produced an essentially scale-free
fluctuation spectrum, though without inflation. This alternative mechanism is
based on the Hollands-Wald concept of a minimum wavelength for the emergence
of quantum fluctuations into the semi-classical universe. A cosmology with
zero active mass does not have a horizon problem, so it does not need inflation
to solve this particular (non) issue. In this picture, the $1^\circ-10^\circ$
fluctuations in the CMB correspond almost exactly to the Planck length at
the Planck time, firmly supporting the view that CMB observations may already 
be probing trans-Planckian physics.
\end{abstract}

\pacno{04.20.Ex, 95.36.+x, 98.80.-k, 98.80.Jk}

\section{Introduction}
In spite of its many successes, standard big-bang cosmology suffers from
several conceptual and physical anomalies and inexplicable observational
puzzles, such as the `horizon' problem, in which the cosmic microwave
background (CMB) temperature is relatively uniform everywhere, even though
causally connected regions at last scattering are much smaller than the
horizon size today. Founded on a combination of classical and quantum
physical principles \cite{Kolb:1990,Linde:1990,Liddle:2000}, the inflationary
paradigm was developed to address these issues
\cite{Kazanas:1980,Starobinsky:1980,Guth:1981,Sato:1981a,Sato:1981b}.
But though the idea of inflation is very flexible,
it has yet to find expression in a comprehensive, self-consistent model
that accounts for all of the observations \cite{Turner:2008}.

Many variations of the inflationary concept exist today
\cite{Linde:1990,Liddle:2000,Riotto:2002,Martin:2005,Bassett:2006,Kinney:2009,Sriramkumar:2009,Baumann:2009},
allowing us to parametrize features in the early universe, but none may
represent a final comprehensive answer if their fundamentally semi-classical
nature is at odds with Planckian (or even pre-Planckian) scale physics. On
the other hand, an inflationary phase may have been associated with a scalar
(`inflaton') field beyond the Planck scale, whose identity could eventually
be revealed via extensions to the Standard Model based on supergravity,
grand unified theories, or even string theory.

Most of the problems arising in standard big-bang cosmology are due to
the inevitable decelerated expansion associated with a radiation or matter
dominated cosmic fluid. Inflation circumvents this deficiency by postulating
the existence of an early phase of accelerated expansion, during which
proper distances grew faster than the gravitational horizon size, which
in some inflationary scenarios actually did not grow at all (or grew
very slowly) during this brief period. Thus, physical distances would have
been pushed beyond the Hubble radius, which potentially solves the horizon
problem.

An additional attractive feature of inflation---central to the subject of
this paper---is that quantum fluctuations of the inflaton field may have
generated density perturbations seeding the formation of large-scale structure
\cite{Mukhanov:1981,Guth:1982,Hawking:1982,Starobinsky:1982}. These fluctuations
would have been stretched on large scales by the brief accelerated expansion
and, in the simplest version of the single inflaton field scenario, would
have become `frozen' after their wavelength exceeded the Hubble radius. And
since inflation must have somehow ended in order for the Universe to
subsequently re-establish its radiation and matter dominated expansion,
the perturbations would have crossed back inside the Hubble radius.

In this picture, fluctuations with ever larger co-moving wavenumbers
crossed the horizon and became frozen at progressively later times, a process
naturally producing nearly scale-invariant spectra. This expectation has
been largely confirmed by measurements of the temperature anisotropies in the
CMB, generating considerable enthusiasm for the inflationary paradigm,
well beyond its early success in apparently resolving other long-standing issues,
such as the aforementioned horizon problem.

Even so, tension continues to grow between the overall expectations of the
inflationary model and some key observations including, and especially, of the
CMB anisotropies. The emergence of greater detail in all-sky maps has revealed
several unexpected features on large scales, first reported by the Cosmic
Background Explorer (COBE) Differential Microwave Radiometer (DMR) collaboration
\cite{Wright:1996}. Disagreement with theory arises from an apparent alignment
of the largest modes of CMB anisotropy, as well as the absence of any angular
correlation at angles greater than $\sim 60^\circ$. The latter is particularly
troublesome because all fluctuations presumably exited the horizon during inflation,
which should have produced a correlation at all angles. These unexpected
features have been explained as possibly due to cosmic variance within
the standard model \cite{Bennett:2013}, but this explanation may not
be completely satisfactory.

An analysis of the differences between the observed angular correlation
function and that predicted by inflation in $\Lambda$CDM \cite{Copi:2009}
has revealed that only $\sim 0.03\%$ of $\Lambda$CDM model CMB skies
have a variance larger than that of the sky observed with the Wilkinson
Microwave Anisotropy Probe (WMAP) \cite{Bennett:2013}. We may simply be
dealing with foreground subtraction issues. But there are indications
that the differences between theory and observations may be due to more
than just randomness. For example, the well-defined shape of the
observed angular correlation function, with a minimum at $\sim 50^\circ$,
is at odds with the expectation that the data points would not have
lined up as they do within the variance window if stochastic processes
were solely to blame. More importantly, the observed angular correlation
goes to zero beyond $\sim 60^\circ$. While variance could have resulted
in a function with a different slope than that predicted by inflation,
it seems unlikely that this randomly generated slope would be close to
zero above $\sim 60^\circ$.

This tension has been exacerbated by the more recent {\it Planck}
results \cite{Ade:2015}. The probability of the {\it Planck} sky
being consistent with inflation in $\Lambda$CDM is $\sim 0.33\%$
for any of the analyzed combinations of maps and masks \cite{Copi:2013},
a trend that has now remained intact through three different
satellite missions (COBE, WMAP, {\it Planck}). The apparent lack
of temperature correlations at large angles is robust and increases
in statistical significance as the quality of the measurements improves,
suggesting that instrumental issues are not the cause. Indeed, if it
turns out that the absence of large-angle correlation is real, this may
be the most significant outcome of the CMB observations, because it
would essentially invalidate any role that inflation might have
played in the universal expansion.

In this paper, we consider an alternative scenario for generating
quantum fluctuations in the early universe, not solely because of
the potential problems facing inflation in accounting for all the
data, but more so because another Friedmann-Robertson-Walker (FRW)
cosmology, known as the $R_{\rm h}=ct$ universe \cite{Melia:2007,MeliaShevchuk:2012}
(see also Ref.~\cite{Melia:2012a} for a non-technical introduction) has been shown
in recent years to account for a broad range of high-precision cosmological
measurements better (in some cases, significantly better) than $\Lambda$CDM,
the current standard model based on inflation.

For example, whereas the inflationary paradigm has trouble explaining the
absence of any angular correlation in the CMB beyond $\sim 60^\circ$, this
characteristic simply results from the size of the gravitational horizon (i.e.,
the Hubble radius) at last scattering in the $R_{\rm h}=ct$ universe
\cite{Melia:2014a}. A rather compelling example of how the predictions
of $\Lambda$CDM and $R_{\rm h}=ct$ differ in their comparisons with the
data is provided by a recent application of the Alcock-Paczy\'nski
test \cite{Alcock:1979}, based on the changing ratio of angular to
spatial/redshift size of (presumed) spherically-symmetric source distributions
with distance, to the most accurate measurements of the baryon acoustic oscillation
(BAO) scale. The use of this diagnostic with newly acquired data on the anisotropic
distribution of the BAO peaks from SDSS-III/BOSS-DR11 at average redshifts
$\langle z\rangle=0.57$ \cite{Anderson:2014} and $\langle z\rangle=2.34$ \cite{Delubac:2015},
disfavors the current concordance ($\Lambda$CDM) model at better than a
$99.34\%$ CL, while the probability that the $R_{\rm  h}=ct$ universe
is consistent with these data is $\sim 0.96$, i.e., essentially one
\cite{Melia:2015a}.

The question concerning how quantum fluctuations were generated in the early
universe is critical to this whole discussion because, whereas $\Lambda$CDM
probably cannot survive without inflation, the $R_{\rm h}=ct$ universe does not
have or need it. This cosmology did not undergo a period of decelerated expansion,
and therefore avoids the horizon problem altogether \cite{Melia:2013a}. So while
this paper is in principle motivated separately by (1) a desire to alleviate the
growing tension between the predictions of inflation and the ever improving
observations, and (2) a need to strengthen the viability of the $R_{\rm h}=ct$
universe by uncovering a mechanism to generate cosmological perturbations
in this model, in reality these two goals overlap considerably. Our principal
task is to determine how and why quantum fluctuations could have grown in
$R_{\rm h}=ct$ without inflation.

Before we begin our development of this mechanism, however, it is worthwhile
considering several earlier attempts at producing quantum fluctuations without
inflation, and how they differ from the proposal we are making here. In their
work, Bengochea et al. \cite{Bengochea:2015} adopted the Hollands-Wand concept,
but focused primarily on the question of how classicalization may actually
occur in such a scenario. This issue, of how a homogeneous and isotropic 
quantum fluctuation is converted into actual inhomogeneities and anisotropies
at the classical scale, is common to all models invoking a quantum origin
for the perturbations, and is yet to be resolved (see also ref.~\cite{Perez:2006}). 
In their analysis, these authors adopted a standard cosmological background 
(other than inflation), whereas in this paper we will focus exclusively on 
the zero active mass equation-of-state associated with the $R_{\rm h}=ct$ 
model. The manner in which the modes are born and subsequently stretch 
and grow is quite different in these two cases. As we shall see, the mode 
wavelength grows as a constant fraction of the Hubble radius, so its 
transition from the Planck domain to the $\sim 1^\circ-10^\circ$ scale 
associated with the CMB is smooth and does not involve multiple steps, 
such as one encounters during inflation, where modes cross and re-cross 
the horizon during their evolution. Nonetheless, this paper will not be 
fully addressing the question of classicalization, which remains a largely 
unresolved problem.

A non-inflationary mechanism for generating the perturbation spectrum
has also been considered in ekpyrotic \cite{Khoury:2001} and cyclic 
\cite{Steinhardt:2002} models. Here too, however, these models have
the common feature that the perturbations originated as quantum fluctuations
which exited and re-entered the horizon during their evolution. Interestingly,
this process occurs for both expanding cosmologies (e.g., in the standard
model) and a contracting universe, with an appropriate alteration to the
evolution in the scale factor $a(t)$ \cite{Steinhardt:2004}. The cyclic
model repeats its periods of expansion and contraction, the latter of
which is identical to the ekpyrotic case. The $R_{\rm h}=ct$ model that
we focus on in this paper is unique, in that this is the only case
in which all proper distances and the Hubble radius expand at the same 
rate. As we shall see, the mechanism for producing a near-scale free
spectrum is therefore simpler, with the added advantage that the observed
scale of fluctuations in the CMB traces back directly to the Planck 
wavelength at the Planck time. None of the other models have this
feature, which provides some justification for the argument that 
perturbations were essentially trans-Plancking in nature.

In \S~II of this paper, we briefly summarize the origin and essential
characteristics of the $R_{\rm h}=ct$ universe, and then discuss the
cosmological dynamics in this model in \S~III. The cosmological perturbations
are introduced in \S~IV, where we describe some of this model's most
significant predictions. We end with our conclusions in \S~V.

\section{The $R_{\rm h}=ct$ Universe}\label{Rhct}
The $R_{\rm h}=ct$ universe is an FRW cosmology in which the underlying
symmetries of the metric, with particular reference to Weyl's postulate
\cite{Weyl:1923}, are used to incorporate the influence of a gravitational
horizon on the expansion dynamics
\cite{Melia:2007,MeliaShevchuk:2012,MeliaAbdelqader:2009}.
The model is based on standard
general relativity (GR), and the Cosmological principle is adopted from
the start, just like any other FRW cosmology, but it equally addresses
the consequences of Weyl's postulate, whose role in shaping the FRW
metric is often ignored.

It is commonly assumed that Weyl's postulate is already incorporated
into all forms of the FRW metric, and is therefore given far less
attention than the Cosmological principle. Simply stated, Weyl's
postulate holds that any proper distance $R(t)$ is a product of
a universal expansion factor $a(t)$ (dependent only on cosmic time $t$)
and an unchanging co-moving radius $r$: $R(t)=a(t)r$. We
conventionally write the FRW metric adopting this coordinate
definition, along with $t$, which is actually the observer's proper
time in his/her free-falling frame.

But its impact is far greater than this. Consider, for instance, the
Misner-Sharp mass $M$---defined in terms of the proper mass density $\rho/c^2$
and proper volume $4\pi R^3/3$ \cite{Misner:1964}---in this spacetime.
In terms of the {\it proper} mass $M$, the gravitational radius of the
Universe is $R_{\rm h}\equiv 2GM/c^2$ \cite{Melia:2007,MeliaAbdelqader:2009},
which actually coincides with the better known Hubble radius $c/H(t)$.
Given its definition, $R_{\rm h}$ (and therefore the Hubble radius)
is a proper distance \cite{MeliaAbdelqader:2009}, so this radius must comply
with Weyl's postulate, the consequence of which is the unique choice
$a(t)=(t/t_0)$ for the expansion factor, where $t_0$ is the current age
of the Universe \cite{MeliaShevchuk:2012}. Those familiar with the properties
of the Schwarzschild or Kerr metrics are not at all surprised by this
constraint, which leads to the result that the gravitational radius must
be receding from us at speed $c$---hence the name `$R_{\rm h}=ct$' for
this model. The Hubble radius was in fact defined to be the distance
at which the Hubble speed equals $c$ even before it was recognized as
another manifestation of the gravitational horizon. In black-hole
spacetimes, a free-falling observer sees the event horizon
approaching them at speed $c$, so this property of $R_{\rm h}=ct$
is quite familiar in the context of standard GR.

One of the principal differences between $R_{\rm h}=ct$ and other
FRW cosmologies, such as $\Lambda$CDM, is how they handle the energy
density $\rho$ and pressure $p$, and their temporal evolution. In
$\Lambda$CDM we routinely start with the constituents in the cosmic
fluid, and assume their equations-of-state, and then solve the dynamics equations
to determine the expansion rate as a function of time. In $R_{\rm h}=ct$,
on the other hand, the symmetries of the FRW metric and the
properties of the gravitational horizon, uniquely specify the spacetime
curvature, and hence the expansion rate, strictly from just the value
of the total energy density $\rho$, without us having to know the specifics
of the constituents themselves. In this  model, the constituents of the
Universe must partition themselves in such a way as to satisfy the
constant expansion rate required by the $R_{\rm h}=ct$ condition.
Insofar as the dynamics is concerned, all that matters is
$\rho$ and the overall equation of state $p\equiv w\rho$. So while
one assumes $\rho=\rho_m+\rho_r+\rho_\Lambda$ in $\Lambda$CDM,
i.e., that the principal constituents are matter, radiation, and
a cosmological constant $\Lambda$, and then infers $w$ from the
equations-of-state assigned to them, in $R_{\rm h}=ct$, it is
the aforementioned symmetries and other constraints from GR
that force the $R_{\rm h}=ct$ universe to have the unique
equation-of-state \cite{Melia:2007,MeliaShevchuk:2012}
\begin{equation}
\rho+3p=0\;.
\end{equation}

The $R_{\rm h}=ct$ cosmology is therefore simple and
elegant, in the sense that observable quantities, such as the
luminosity distance $d_L$ and the redshift dependence of the
Hubble constant $H$, take on analytic forms:
\begin{equation}
d_L=R_h(t_0)(1+z)\ln(1+z)\;,
\end{equation}
and
\begin{equation}
H(z)=H_0(1+z)\;,
\end{equation}
where $z$ is the redshift, $R_h(t)=c/H(t)$, and $H_0$ is the value
of the Hubble constant today. Relations such as these have been tested
using a broad range of measurements, such as the BAO observations
described above, and have thus far accounted for the data better
than their counterparts in $\Lambda$CDM
\cite{Melia:2014a,Melia:2013b,MeliaMaier:2013,Wei:2013,Melia:2014b,Wei:2014a,Melia:2014c,Wei:2014b,Melia:2015b,Melia:2015c,Wei:2015a,Wei:2015b,Wei:2015c,Wei:2015d,Melia:2015d,Melia:2015e}.

Nonetheless, with its empirical approach, $\Lambda$CDM has done
remarkably well as a reasonable approximation to $R_{\rm h}=ct$
in restricted redshift ranges. For example, using the ansatz
$\rho=\rho_m+\rho_r+\rho_\Lambda$ to fit the data, one finds
that the parameters in the standard model must have quite specific
values, such as $\Omega_m\equiv \rho_m/\rho_c\sim 0.27$, where
$\rho_c$ is the critical density \cite{Melia:2015f}. However, with
these parameters, $\Lambda$CDM then requires $R_{\rm h}(t_0)\approx ct_0$
today, which is in fact the baseline constraint in the $R_{\rm h}=ct$
model. One concludes from this that the optimized $\Lambda$CDM
cosmology describes a universal expansion equal to what it would
have been with $R_{\rm h}=ct$ all along. And many other
indicators support the view that using $\Lambda$CDM to fit
the data therefore produces a cosmology almost, but not entirely
identical, to $R_{\rm h}=ct$, in spite of the fact that with
its many free parameters, $\Lambda$CDM could have had an entirely
diverse set of expansion histories.

\section{Cosmological Dynamics}\label{dynamics}
\subsection{Field Equations}
The Friedmann-Robertson-Walker metric is conventionally written
in terms of the comoving coordinates $(t,r,\theta,\phi)$, and
takes the form
\begin{equation}
ds^2=dt^2-a(t)^2\left[{dr^2\over 1-Kr^2}+r^2\left(d\theta^2
+\sin^2\theta\,d\phi^2\right)\right]\;,
\end{equation}
where $K$ is the spatial curvature constant. In $R_{\rm h}=ct$,
$K$ must be zero in order for the gravitational radius to coincide
with the Hubble radius, which the data (interpreted in the context
of $\Lambda$CDM) all seem to be confirming as well. We will therefore
henceforth set $K=0$ in all our derivations.

The dynamical equations for this background FRW metric are obtained
from Einstein's equations,
\begin{equation}
G_{\alpha\beta}\equiv \mathcal{R}_{\alpha\beta}-{1\over 2}g_{\alpha\beta}\mathcal{R}=
-{8\pi G}T_{\alpha\beta}\;,
\end{equation}
where $g_{\alpha\beta}$ are the metric coefficients, and $\mathcal{R}_{\alpha\beta}$
and $\mathcal{R}$ are the Ricci tensor and scalar, respectively:
\begin{equation}
H^2\equiv \left({\dot{a}\over a}\right)^2={8\pi G\over 3}\rho\;,
\end{equation}
and
\begin{equation}
{\ddot{a}\over a}=-{4\pi G\over 3}\left(\rho+3p\right)\;,
\end{equation}
where a dot denotes a derivative with respect to $t$, and $\rho$ and
$p$ are, of course, the proper energy density and pressure in the
co-moving frame. Throughout this paper, we work with natural units,
in which $\hbar=c=1$. The continuity equation for the (perfect fluid) 
energy-momentum tensor,
\begin{equation}
T_{\alpha\beta}=\left(\rho+p\right)u_\alpha u_\beta-pg_{\alpha\beta}\;,
\end{equation}
in terms of the four-velocity $u_\alpha$, yields a third (though not independent)
equation, expressing the (local) conservation of energy:
\begin{equation}
\dot{\rho}=-3H(\rho+p)\;.
\end{equation}

\subsection{The Numen Field}
In the inflationary model, one assumes the existence of scalar inflaton
fields that dominate $\rho$ in the cosmic fluid
prior to the onset of leptogenesis and baryogenesis, with properties
that lead to an exponential solution for $a(t)$ in Equations~(6) and (7),
thus heralding a very brief period of de Sitter expansion \cite{deSitter:1917}.
It is not difficult to imagine such fields influencing cosmological dynamics
in the very early universe. For example, a grand unified theory based
on the group SO(10) implies the existence of $\sim 100$ Higgs fields,
and supersymmetry relies on the existence of many superpartners
\cite{Lyth:1999}. In addition, string theory has dynamical moduli fields
associated with the geometrical characteristics of compactified
dimensions \cite{Lidsey:2000}.

Though the $R_{\rm h}=ct$ universe does not need or have inflation, and
therefore does not require the presence of an `inflaton' field, we will
nonetheless assume that at least one scalar field dominated
the cosmological dynamics at the very beginning. But unlike the situation
with the standard model, the expansion factor $a(t)$ in the $R_{\rm h}=ct$
universe must always be proportional to $t$, so the expansion in this
cosmology is never inflated. To clearly distinguish the hypothesized scalar
field $\phi$ associated with the expansion in this model from those generally
categorized as `inflaton' fields, we will therefore refer to it informally
as the `numen' field, giving rise to the earliest manifestation of substance
in the nascent universe, with an equation-of-state $\rho_\phi+3p_\phi=0$
and, as we shall see very shortly, whose quantum fluctuations might have
seeded the subsequent formation of large-scale structure.

A crucial difference between the $R_{\rm  h}=ct$ universe and other FRW
cosmologies is that the cosmic fluid in the former has strictly `zero active
mass,' meaning that $\rho+3p=0$, so the expansion proceeds without any
net gravitational influence (after all, this is why $\dot{a}$=constant).
This suggests that coupling the numen field non-minimally to gravity
using the simple prescription $\xi \mathcal{R}\phi^2/2$
\cite{Tsujikawa:2000,delCampo:2015} (where $\xi$ is a dimensionless
coupling constant) in the Lagrangian density may not be as relevant
here as it is in the inflaton case (though not impossible, of course).
We will rely on the simplest assumption we can make, i.e., that the
background dynamics is dominated by a single homogeneous minimally-coupled
scalar field with action
\begin{equation}
S=\int d^4x\sqrt{-g}\;\mathcal{L}(\phi,\partial_\mu\phi)\;,
\end{equation}
where $\sqrt{-g}=a^3(t)$ for the metric in Equation~(4), and
the Lagrangian density is given as
\begin{equation}
\mathcal{L} = {m_{\rm P}^2\over 16\pi}\mathcal{R}+{1\over 2}\partial^\mu\phi
\partial_\mu\phi-V(\phi)\;,
\end{equation}
where $m_{\rm P}\equiv G^{-1/2}$ is the Planck mass.
As it turns out, the potential $V(\phi)$ for the numen field $\phi$
is unique in $R_{\rm h}=ct$, and we shall derive it very shortly.

Since the (background) field is homogeneous, we can ignore spatial
gradients, and so the corresponding energy density $\rho_\phi$
($=T_{00}$) and pressure $p_\phi$ ($=T_{ii}$) are given simply as
\begin{equation}
\rho_\phi={1\over 2}{\dot{\phi}}^2+V(\phi)\;,
\end{equation}
and
\begin{equation}
p_\phi={1\over 2}{\dot{\phi}}^2-V(\phi)\;.
\end{equation}
The zero active mass condition therefore immediately constrains the potential
to have the unique form
\begin{equation}
V(\phi)={{\dot{\phi}}^2}\;,
\end{equation}
and the energy conservation Equation~(9) gives
\begin{equation}
\ddot{\phi}+3H\dot{\phi}+{\partial V\over\partial \phi}=0\;,
\end{equation}
the usual Klein-Gordon equation.

The Friedmann Equation~(6) similarly reduces to a very simple form,
\begin{equation}
H^2={4\pi \over m_{\rm P}^2}V(\phi)\;,
\end{equation}
and combining this with Equation~(14) then allows us to find an exact
solution for the numen field:
\begin{equation}
\phi(t)-\phi(t_i)={m_{\rm P}\over \sqrt{4\pi}}\ln\left({t\over t_i}\right)\;,
\end{equation}
where $t_i$ is some fiducial time at which the field has an
amplitude $\phi(t_i)$. Returning to Equation~(14), we therefore also
have an exact---and unique---solution for the numen potential:
\begin{equation}
V(\phi)=V_0\,\exp\left\{-{2\sqrt{4\pi}\over m_{\rm P}}\,\phi\right\}
\end{equation}
where, for convenience, we have defined the constant
\begin{equation}
V_0\equiv {m_{\rm P}^2\over 4\pi \, t_i^2}\exp
\left\{{2\sqrt{4\pi}\over m_{\rm P}}\phi(t_i)\right\}\;.
\end{equation}

It is interesting to note that before settling on de Sitter (or quasi de
Sitter) expansion for the inflationary paradigm, there were attempts in
the 1980's to consider minimally coupled inflaton fields with an exponential
potential
\begin{equation}
V_p(\phi)=V_0\,\exp\bigg\{\pm {2\over p}{\sqrt{4\pi}\over m_{\rm P}}\,\phi\bigg\}\;,
\end{equation}
as a means of producing so-called power-law inflation (PLI) with $p>1$
\cite{Abbott:1984,Lucchin:1985,Barrow:1987,Liddle:1989}. During PLI, the scale
factor has the time dependence
\begin{equation}
a(t)\propto t^p\;.
\end{equation}
Again, the intention with these was to circumvent the problems arising
from deceleration in standard big bang cosmology. The numen-field potential
(Equation~18) is clearly a special member of this class, though with
$p=1$ it does not inflate. Exponential potentials such as these are
generally motivated in the context of Kaluza-Klein cosmologies \cite{Lucchin:1985},
and arise in string theories, supergravity, and actually any theory
based on a conformal transformation to the Einstein frame.

A difficulty commonly encountered with inflaton models is that they lack
an exit mechanism for a decelerating (radiation or matter dominated) phase
to succeed inflation. In contrast, the expansion rate is always
constant in $R_{\rm h}=ct$, so no dynamical transition is required
as the numen field decays into radiation and other particles in
the standard model (via channels yet to be determined).

\section{Cosmological Perturbations}\label{perturbations}
We have learned from inflationary cosmology that to properly interpret
aniso\-tropies in the CMB, one needs a description of the fluctuations
characterized by several observables. These include: (1) the scalar
spectral index $n_s$, (2) the spectral index $n_T$ of the tensor
perturbations and (where possible) (3) the tensor-to-scalar ratio $r$,
giving the ratio of tensor to scalar amplitudes \cite{Starobinsky:1979,Starobinsky:1985}.
The Wilkinson Microwave Anisotropy Probe (WMAP) \cite{Bennett:2013} and {\it Planck}
\cite{Ade:2015} have placed strong bounds on at least some of these parameters:
$n_s=0.968\pm0.006$ (corresponding to essentially a scale-free spectrum)
and $r<0.11$ at $95\%$ CL.

Let us now consider small perturbations about the homogeneous numen field $\phi_0(t)$,
\begin{equation}
\phi(t,\vec{x}) = \phi_0(t)+\delta\phi(t,\vec{x})\;,
\end{equation}
keeping only terms to first order in $\delta\phi$. The inhomogeneity implied
by these fluctuations requires that we also include metric perturbations
about the spatially flat FRW background metric, which can be conveniently
split into scalar, vector, and tensor components, depending on how they
transform on spatial hypersurfaces.

By now, it is well known that vector perturbations have no lasting influence
for scalar fields. The perturbed FRW spacetime for the remaining linearized scalar
and tensor fluctuations is therefore described by the line element
\cite{Bassett:2006,Bardeen:1980,Kodama:1984,Mukhanov:1992}
\begin{eqnarray}
ds^2 &=& (1+2A)\,dt^2-2a(t)(\partial_iB)\,dt\,dx^i-\nonumber\\ 
&\null& a^2(t)\left[(1-2\psi)\delta_{ij}+2(\partial_i
\partial_jE)+h_{ij}\right]\,dx^i\,dx^j\,,\quad
\end{eqnarray}
where indices $i$ and $j$ denote spatial coordinates, and $A$, $B$, $\psi$
and $E$ describe the scalar degree of metric perturbations, while $h_{ij}$
represent the tensor perturbations. This form follows the notation of
Ref.~\cite{Mukhanov:1992}, aside from the use of the symbol $A$ instead of
$\phi$ inside the lapse function (since we are reserving this symbol to
represent the scalar field in this paper).

The Einstein equations for the scalar and tensor parts decouple to linear order,
but the form of the scalar equation is gauge dependent. However, one can identify
a variety of gauge-independent combinations of the scalar perturbations from
within certain coordinate systems. For example, in the comoving frame, the curvature
perturbation $\Theta$ on hypersurfaces orthogonal to comoving worldlines may
be defined \cite{Bardeen:1980} as a gauge invariant combination of the metric
perturbation $\psi$ and the scalar field perturbation $\delta\phi$:
\begin{equation}
\Theta\equiv \psi+\left({H\over \dot{\phi}}\right)\,\delta\phi\;.
\end{equation}
\subsection{Scalar Perturbations}
Expanding $\Theta$ in Fourier modes,
\begin{equation}
\Theta(t,\vec{x})=\int {d^3\vec{k}\over (2\pi)^{3/2}}\,\Theta_k(t)e^{i\vec{k}\cdot\vec{x}}\;,
\end{equation}
where $k$ is the comoving wavenumber, and using the linearized version of Einstein's
Equations~(5) with the linearized metric in Equation~(23), one arrives at the perturbed
equation of motion
\begin{equation}
\Theta_k^{\prime\prime}+2\left({z^\prime\over z}\right)\Theta_k^\prime+k^2\Theta_k=0\;,
\end{equation}
where overprime now denotes a derivative with respect to conformal time $d\tau\equiv {dt/a(t)}$.
In the $R_{\rm h}=ct$ universe, we have $a(t)=(t/t_0)$, where $t_0$ is a fixed time
usually taken to be the present age of the Universe, so that $a(t_0)=1$. Therefore
\begin{equation}
\tau(t) = \tau_i+t_0\ln\left({t\over t_i}\right)\;,
\end{equation}
where $\tau_i$ is the conformal time at some fiducial cosmic time $t_i$. To simplify the
notation, we will define the zero of conformal time to be at $t_0$, so throughout this paper
we will employ the relation
\begin{equation}
\tau(t) = t_0\ln a(t)\;.
\end{equation}

The quantity $z$ in Equation~(26) is defined by the expression
\begin{equation}
z\equiv {a(t)(\rho_\phi+p_\phi)^{1/2}\over H}\;.
\end{equation}
Using Equations~(6), (12) and (13) for the numen field, $z$ reduces to the much
simpler form
\begin{equation}
z={m_{\rm P}\over\sqrt{4\pi}}a(t)\;,
\end{equation}
and so $z^\prime/z=1/t_0$ and $z^{\prime\prime}/z=1/t_0^2$.

The quantity $z^\prime/z$ typically depends on the background dynamics,
so one often conveniently rewrites Equation~(26) in terms of the so-called
Mukhanov-Sasaki variable $u_k\equiv z\,\Theta_k$ \cite{Sasaki:1986,Mukhanov:1988}.
It is not really necessary to do this here since $z^\prime/z$ and $z^{\prime\prime}/z$
are actually constant for the numen field, but we will take this step anyway just
to make it easier to directly compare the differences between our solution and that
pertaining to conventional inflaton fields. With this change of variable,
the equation governing the curvature perturbation now becomes
\begin{equation}
u_k^{\prime\prime}+\alpha_k^2u_k=0\;,
\end{equation}
where
\begin{equation}
\alpha_k\equiv {1\over t_0}\sqrt{\left(2\pi R_{\rm h}\over \lambda_k\right)^2-1}
={1\over t_0}\sqrt{\left(k R_{\rm h}\over a\right)^2-1}\;,
\end{equation}
and $\lambda_k\equiv 2\pi a/k$ is the proper wavelength corresponding to comoving wavenumber $k$.

This expression for the `frequency' $\alpha_k$ is critical to understanding
the nature of quantum fluctuations in the numen field. Its most distinct departure
from inflaton fields is that both the gravitational radius $R_{\rm h}=t$, and the
proper wavelength $\lambda_k\sim a(t)$, scale with time in exactly the same way,
so the ratio $R_{\rm h}/\lambda_k$ or, equivalently, $kR_{\rm h}/a$, is constant.
In this cosmology there is no crossing of wave modes back and forth across the
horizon. In fact, once the wavelength of a mode is established when it emerges
into the semi-classical universe, it remains a fixed fraction of the Hubble 
radius while both expand with time. And notice that all modes $u_k$ with a 
wavelength smaller than the horizon, $R_{\rm h} >\lambda_k/2\pi$, oscillate, 
while those with a super-horizon wavelength do not. (In this particular regard, 
the numen and inflaton fields behave similarly.) The analytic solution to 
Equation~(31) may be written as follows:
\begin{equation}
u_k(\tau) = \left\{ \begin{array}{ll}
         B(k)\,e^{\pm i\alpha_k\tau} & \mbox{($2\pi R_{\rm h}>\lambda_k$)} \\
         B(k)\,e^{\pm |\alpha_k|\tau} & \mbox{($2\pi R_{\rm h}<\lambda_k$)}\end{array} \right. 
\end{equation}

The amplitude $B(k)$ is fixed by an appropriate choice of vacuum,
related to how these modes are ``born." Quantum fluctuations of the
inflaton field are created with a wavelength much smaller than the
horizon, so they behave at first like an ordinary harmonic oscillator
(similar to the oscillatory solution in Equation~33). But during inflation,
$H$ is essentially constant, so $\lambda_k$ overtakes the Hubble
radius $R_{\rm h}=1/H$ and becomes much larger, and the mode becomes
an overdamped oscillator, with an amplitude that approaches a constant
value, a process often referred to as ``freezing." Once inflation has
ended, the Hubble radius resumes its rate of growth and overtakes
$\lambda_k$, which is said to then ``re-enter" the horizon.

In a time-independent spacetime, a preferred set of mode functions, and therefore
an unambiguous physical vacuum, may be defined by minimizing the expectation
value of the Hamiltonian. In Minkowski space, this means taking the positive
frequency mode $u_k\sim e^{-ick\tau}$, i.e., the minimal excitation state,
and setting $B(k)=1/\sqrt{2k}$ \cite{Bassett:2006}. This prescription,
however, does not usually generalize straightforwardly to time-dependent
spacetimes, but this vacuum ambiguity can still be resolved in inflationary
models by arguing that in the remote past all observable modes had a
wavelength much smaller than the horizon, and were therefore not affected by
gravity, so their frequencies were essentially time-independent. They
therefore behaved as they would in Minkowski space. This approach defines
the preferred set of mode functions and a unique physical vacuum known
as the Bunch-Davies vacuum \cite{Bunch:1978}. The amplitude $B(k)$ for
super-horizon modes is then evaluated from the Bunch-Davies normalization
by equating the amplitudes before and after freezing. And because the value
of $a(t)$ at which the freezing occurs is proportional to $k$ (via the condition
$k/a\sim 1/R_{\rm h}$), this process results in a scale-free spectrum (see
below), consistent with the observed anisotropies in the CMB, considered to
be one of the strongest factors in favor of the inflationary paradigm.

However, some have questioned the fundamental basis for this picture because
in many models of inflation, the de Sitter phase lasted so long that the
inflaton modes responsible for the creation of large-scale structure
would have been born with wavelengths much shorter than the Planck scale
(and therefore well before the Planck time), where the use of semi-classical
physics is uncertain. This ``trans-Planckian issue" \cite{Martin:2001}
revolves around the question of whether the semi-classical description
of our Universe breaks down prior to the Planck time, set by the condition
that the Compton wavelength $\lambda_{\rm C} \equiv 2\pi/m$ of a mass $m$
be equal to its Schwarzschild radius $R_{\rm h}\equiv 2Gm$. Current
physics may have to be modified on spatial scales smaller than the resulting
Planck length $\lambda_{\rm P}\equiv\lambda_{\rm C}(m)$ at the specific
value of $m$ where this equality is reached:
\begin{equation}
\lambda_{\rm P}=\sqrt{4\pi G}\;.
\end{equation}
The corresponding Planck time $t_{\rm P}$ is simply this Planck length
divided by the speed of light $c$. Numerically, we have $\lambda_{\rm P}\approx 
5.7\times 10^{-33}$ cm and $t_{\rm P}\approx 1.9\times 10^{-43}$ s.
These definitions actually make more sense for the $R_{\rm h}=ct$
universe than they do for the standard inflationary model,
because the gravitational radius $R_{\rm h}$ is in fact equal to $t$,
so the Planck time is simply the age of the Universe when the Hubble
radius equaled the Planck scale (i.e., $R_{\rm h}=\lambda_{\rm P}$).

Let us now track the mode growth associated with the CMB anisotropies
in the $R_{\rm h}=ct$ universe back to these earlier times and see
how they are related to $\lambda_{\rm P}$ and $t_{\rm P}$ in this cosmology.
The CMB spectrum has features ranging from sub-degree scales to tens of
degrees. The Sachs-Wolfe effect \cite{Sachs:1967}, responsible for
coupling the metric fluctuations with the primordial perturbations,
contributes to temperature anisotropies on all scales, but tends to
dominate at angles $\gtrsim 1^\circ-10^\circ$. On sub-degree scales, the
spectral peaks are primarily dependent on the pressure and density
variations associated with baryon acoustic oscillations. The
characteristic CMB scale representing the effects of scalar/metric fluctuations
therefore appears to be $\sim 1^\circ-10^\circ$.

In the $R_{\rm h}=ct$ cosmology, the angular-diameter distance is given
as \cite{Melia:2007,MeliaShevchuk:2012,Wei:2015a}
\begin{equation}
d_A={R_{\rm h}(t_0)\over (1+z)}\ln(1+z)\;.
\end{equation}
Therefore, a $\theta$-fluctuation at redshift $z_{\rm CMB}$ corresponds
to a proper wavelength
\begin{equation}
\lambda^{\theta}(z_{\rm CMB})=2\pi\left({\theta\over 360^\circ}\right){R_{\rm h}(t_0)
\over (1+z_{\rm CMB})}\ln(1+z_{\rm CMB})\;.
\end{equation}
And with $a(t)=t/t_0$, it is straightforward to see that at the Planck
redshift, $z_{\rm P}\equiv t_0/t_{\rm P}-1$, the numen-field mode
responsible for this anisotropy had a corresponding wavelength
$\lambda^{\theta}(z_{\rm P})$ given by the expression
\begin{equation}
{\lambda^{\theta}(z_{\rm P})\over\lambda_{\rm P}}\approx \left({\theta\over 57^\circ}\right)
\ln(1+z_{\rm CMB})\;.
\end{equation}
A precise estimate for $z_{\rm CMB}$ does not exist yet for the $R_{\rm h}=ct$ universe, but
this uncertainty has negligible impact on the use of Equation~(37) because the behavior of
$d_A$ with redshift in this cosmology renders $\lambda^{\theta}(z_{\rm P})/\lambda_{\rm P}$
only weakly dependent on the redshift at last scattering. This may be seen in Table~1,
where we quote the ratio $\lambda^{\theta}(z_{\rm P})/\lambda_{\rm P}$ for two angles
($\theta=1^\circ$ and $10^\circ$) and a very broad range of CMB redshifts. Clearly,
the numen-field fluctuations producing the CMB anisotropies had a size comparable
to the Planck scale were we to trace them back to the Planck time. The significance
of this feature should not be underestimated. In the $R_{\rm h}=ct$ cosmology, the
Universe underwent an expansion by over $60$ orders of magnitude in $a(t)$ between
$z_{\rm P}$ and $z_{\rm CMB}$. Yet in this model the observed scale characterizing
the CMB anisotropies tracks back directly to the Planck length at $t_{\rm P}$, in
contrast to standard inflationary cosmology in which the CMB fluctuations have no
obvious connection to the Planck scale. It would be a remarkable coincidence for
$\lambda^{10^\circ}(z_{\rm P}) \sim \lambda_{\rm P}$ if these two scales were not
related dynamically in some way.

\begin{table}[h]
\caption{Perturbation wavelength at $t_{\rm P}$ producing a $\sim$$1^\circ$-$10^\circ$ \\ 
fluctuation in the CMB}
 \centering
  \begin{tabular}{rcc}
&& \\
    \hline
\hline
$z_{\rm CMB}$ & $\lambda^{1^\circ}(z_{\rm P})/\lambda_{\rm P}$ &
$\lambda^{10^\circ}(z_{\rm P})/\lambda_{\rm P}$ \\
\hline
500 & $0.11$ & $1.1$ \\
1,000 & $0.12$ & $1.2$ \\
10,000 & $0.16$ & $1.6$ \\
\hline\hline
  \end{tabular}
\end{table}
\vskip 0.2in
In the context of $R_{\rm h}=ct$, it is therefore quite natural---perhaps even
required---for us to view the modes as having emerged into the semi-classical
Universe starting at the Planck scale $\lambda_{\rm P}$. Such an idea---that modes 
may have been born at a specific physical scale---has already been considered by 
several other authors, particularly Hollands and Wald \cite{Hollands:2002}, who 
focused on the question of where and when a semi-classical description of our 
Universe may be valid. Their context was different from ours, and it was not
clear why the physical scale they introduced (which they called $l_0$) ought
to somehow be related to $\lambda_{\rm P}$. They found that to match the
observed fluctuation amplitude in the CMB, they needed $l_0$ to be five orders 
of magnitude larger than the Planck length. As they noted, however, and as
we shall see below, the Hollands-Wald concept works in a way that makes this
ratio essentially independent of the behavior of $a(t)$, so we will also
conclude that although the numen-field fluctuations might have begun across
the Planck region, their emergence into the semi-classical universe could 
not have been completed on a scale length shorter than $\lambda_0\sim 
10^{5}\lambda_{\rm P}$. 

The Hollands-Wald concept for how quantum fluctuations are born in
this context is based on the assumption that semi-classical physics
applies (at least in some rough sense) to phenomena on spatial scales
larger than this fundamental length $\lambda_0$, so that modes effectively
emerge only when their proper wavelength equals $\lambda_0$. (Note, however,
that the idea of modes being created when their wavelength is at a given
spatial scale is actually not unique. Some previous arguments supporting
this concept may be found in Refs.~\cite{Brandenberger:2002,Hassan:2003}.)
In this view, it makes sense to talk about a classical spacetime metric and 
quantum fields at times earlier than the Planck time $t_{\rm P}$, but only if
this is done with a restriction to phenomena based solely on spatial scales
larger than $\lambda_0$. In this picture, $k$-modes may be created at different 
times rather than all at once, though in a sequence based on the relationship 
between $\lambda_k$ and $\lambda_0$.

\begin{figure} 
\begin{center}
\includegraphics[scale=0.65]{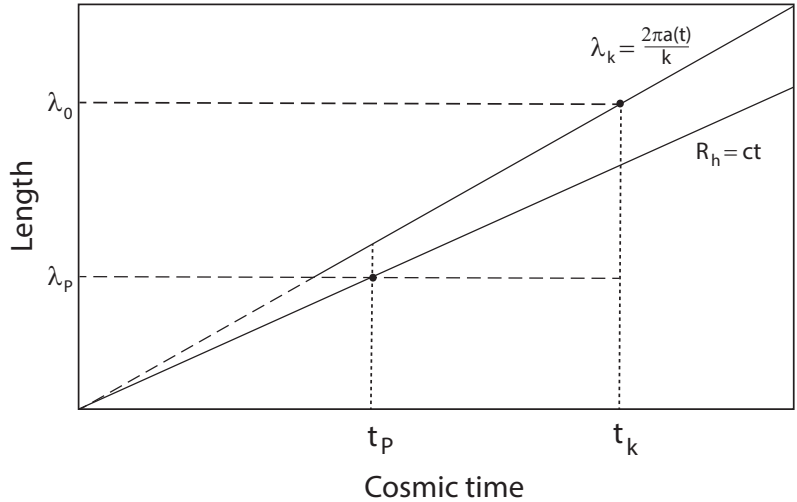}
\end{center}
\caption{Schematic diagram illustrating a $k$-mode born across the
Planck region and emerging into the semi-classical universe when
its wavelength $\lambda_k$ equals the scale $\lambda_0$. The vertical 
axis shows proper distances as a function of cosmic time $t$ 
(increasing to the right on the horizontal axis). The time of birth 
$t_k$ is defined by the condition $\lambda_0=2\pi a(t_k)/k$. In the 
$R_{\rm h}=ct$ cosmology, the ratio $\lambda_k/R_h$ is constant 
for all time $t$.}
\end{figure}

In the $R_{\rm h}=ct$ cosmology, we have several good reasons for adopting
Hollands and Wald's central idea. Chief among them is the empirical evidence
described above, which supports the conclusion that $10^\circ$-anisotropies in
the CMB would have had a size $\sim\lambda_{\rm P}$ at $\sim t_{\rm P}$
(see Table 1). Second, the very notion of a Planck length rests on the physical
limitations imposed on the localization of a defined mass by its Compton
wavelength $\lambda_{\rm C}$. Since the gravitational radius $R_{\rm h}(t)$
defines the maximum size of any causally connected region at time $t$ in this
model \cite{Bikwa:2012,Melia:2012b,Melia:2013c}, only proper masses smaller than
$\sim R_{\rm h}/2G$ have any physical meaning. So quantum mechanical
``fuzziness" extends over a scale $\sim \lambda_{\rm C}(t)=\lambda_{\rm P}^2/t$, 
even bigger than $\lambda_{\rm P}$ at $t<t_{\rm P}$.

As is well known from our experience with fluctuations in the inflaton
field, and as evident in the form of the frequency $\alpha_k$ in Equation~(32), modes
with $\lambda_k<2\pi R_{\rm h}$ oscillate, while those with $\lambda_k>2\pi R_{\rm h}$
are effectively ``frozen" on super-horizon scales (see discussion in \S~1). And as
one finds during inflation, the amplitude of the oscillating metric perturbation
modes would have decayed too quickly for them to be observationally relevant at 
time $t_{\rm cmb}$. So we follow Hollands \& Wald (2002) in identifying the 
Sachs-Wolfe perturbations in the CMB with those trans-Planckian fluctuations 
with super-horizon wavelengths at the time they were born. Figure 1 shows the
key scales relevant to this hypothesis, including the Planck wavelength
$\lambda_{\rm P}=R_{\rm h }(t_{\rm P})$, and the wavelength $\lambda_k=
2\pi a(t)/k$ of mode $k$. We emphasize again the key difference between
the numen-field and inflaton fluctuations, in that the ratio $\lambda_k/R_{\rm h}$
is constant for the former, while it first increases and then decreases during 
inflation.

For these reasons, we adopt the view that all the $k$-modes of
interest in the CMB satisfy the condition
\begin{equation}
k={2\pi\over\epsilon_k\,\lambda_{\rm P}}a(t_{\rm P})\;,
\end{equation}
where $\epsilon_k\gtrsim 1$. They emerge into the semi-classical
universe when their wavelength equals the scale $\lambda_0$, so that
(with $a(t)=t/t_0$)
\begin{equation}
t_k=k{\lambda_0t_0\over 2\pi}\;.
\end{equation}
An alternative way to write this is
\begin{equation}
t_k={\lambda_0\over\epsilon_k\,\lambda_{\rm P}}t_{\rm P}\;.
\end{equation}
We now write Equation~(32) using the approximate expansion 
\begin{equation}
\alpha_k\approx k\left(1-{1\over (kt_0)^2}\right)\;,
\end{equation}
so that with the definition of $z$ in Equations~(29) and (30), 
the metric perturbation $\Theta_k$ has an amplitude frozen at
the birth time $t_k$, given by the expression (cf. ref.~\cite{Hollands:2002})
\begin{eqnarray}
\left|\Theta_k\right|^2 &=& \left.\left({H\over\dot{\phi}}\right)^2{1\over 2 a(t)^2\alpha_k}
\right|_{t_k}\nonumber\\
&\approx&{2\pi^2\over k^3}{\lambda_{\rm P}^2\over \lambda_0^2}\left(1-
{1\over 2(kt_0)^2}\right)^{-1}\;.
\end{eqnarray}

The power spectrum for these curvature perturbations is given by the $k$-space
weighted contribution of modes \cite{Liddle:2000,Bassett:2006,Lidsey:1997,Peiris:2003},
commonly written as
\begin{equation}
\mathcal{P}_{\Theta}(k) \equiv {k^3\over 2\pi^2}|\Theta_k|^2\;.
\end{equation}
We can therefore confirm that the spectrum of numen scalar curvature
perturbations is almost scale free, i.e., $\mathcal{P}_\Theta(k)\sim k^0$,
as seen in the CMB fluctuations, but not exactly. The common way to
quantify the deviation from scale-invariance is via the scalar spectral
index, $n_s$, defined according to
\begin{equation}
n_s-1\equiv {d\,\ln \mathcal{P}_{\Theta}(k)\over d\,\ln k}\;.
\end{equation}
From Equations~(42-44), we find that
\begin{equation}
n_s\sim 1-{2\over 2(kt_0)^2-1}\;.
\end{equation}
As we noted earlier in this section, the observed index is 
$n_s=0.968\pm0.006$ \cite{Ade:2015}, suggesting that the actual power 
spectrum is only approximately scale free. The implication of this 
measurement for the numen-field fluctuations is that the slight
deviation from a pure scale-free spectrum appears to be due to
the difference $k-\alpha_k$ in Equation~(32). Furthermore, as was
the case for Hollands and Wald \cite{Hollands:2002}, we find that
the correct amplitude of the fluctuations in the CMB is produced
if we choose $\lambda_0$ to be of order $10^5\,\lambda_{\rm P}$,
the grand unification scale.

Of course, the value of the ratio in Equation~(45)---and therefore of the 
inferred scalar spectral index $n_s$---depends on the wavenumber $k$. So the 
numen fluctuation spectrum will have a weakly running spectral index. Though 
perhaps not as reliable as the index $n_s$ itself, {\it Planck} constrained 
its scale dependence $dn_s/d\ln k$ to have the value $-0.003\pm0.007$, possibly 
negative, though consistent with zero (or even slightly positive). From 
Equation~(45), we find that the numen field has 
\begin{equation}
{dn_s\over d\ln k}\sim {4(kt_0)^2\over [2(kt_0)^2-1]^2}\;,
\end{equation}
a very small, though positive number. Future work will tell whether this 
difference is meaningful, or whether it is simply due to the fact that the 
analysis reported by {\it Planck} was carried out solely in the context of 
$\Lambda$CDM.

\subsection{Tensor Perturbations}
The tensor perturbations $h_{ij}$ are transverse ($\partial^i h_{ij}=0$)
and trace-free ($\delta^{ij}h_{ij}=0$) and are automatically independent
of coordinate gauge transformations. These represent gravitational waves
evolving independently of linear matter perturbations, and are typically
decomposed into eigenmodes $e_{ij}$ of the spatial Laplacian operator
with comoving wavenumber $k$ and scalar amplitude $h(t)$, such that
\def\cross{\hbox{$\times$}}
\begin{equation}
h_{ij}(t,\vec{x})=h(t)\,e_{ij}^{(+,\cross)}(\vec{x})\;,
\end{equation}
with two independent polarization states $+$ and $\cross$.

In addition to decoupling completely from scalar perturbations
and not providing any backreaction to the metric, gravity waves
also satisfy sourceless equations when the energy-momentum tensor
is diagonal, like that in Equation~(8). With the definition
$v_k(t)\equiv (m_{\rm P}/\sqrt{32\pi})a h_k(t)$ for the Fourier components
$h_k(t)$ of $h$ (based on Equation~47 and our definition of the Planck
mass $m_{\rm P}=G^{-1/2}$), it is easy to see that the mode
equation for the tensor perturbations (analogous to Equation~31) is
\begin{equation} 
v_k^{\prime\prime}+\alpha_k^2v_k=0\;,
\end{equation}
with the same frequency $\alpha_k$ defined in Equation~(32).

The fields $v_k$ and $u_k$ are considered to have similar attributes,
notably, that both are canonically normalized, and that both `freeze' when
their wavelengths exceed the horizon scale. Therefore, we can immediately
write down the expression equivalent to Equation~(42) for the amplitude
of $h_k$:
\begin{equation}
|h_k|^2 = {32\pi\over m_{\rm P}^2}\left.{1\over 2a^2\,\alpha_k}\right|_{\lambda_0^T}\;,
\end{equation}
where $\lambda_0^T$ is the scale---analogous to $\lambda_0$ for the scalar
perturbations---at which the tensor modes emerge into the
semi-classical universe. And therefore since there are two independent
tensor polarization states, the tensor power spectrum (analogous to
Equation~43) is
\begin{eqnarray}
\mathcal{P}_T(k) &\equiv& {2k^3\over 2\pi^2}|h_k|^2\nonumber\\
&=&16{\lambda_{\rm P}^2\over \lambda_0^2}\left(1-{1\over 2(kt_0)^2}\right)^{-1}\;.
\end{eqnarray}

We are now in a position to examine the third observational
signature typically associated with the idea of a quantum-fluctuation 
origin for the anisotropies in the CMB---the ratio of tensor to
scalar power, which we may write as follows:
\begin{equation}
r\equiv{\mathcal{P}_T(k)\over \mathcal{P}_{\Theta}(k)}=
16\left({\lambda_0\over\lambda_0^T}\right)^2\;.
\end{equation}
Those familiar with the inflationary scenario will recognize that
this expression is very similar to the result associated with an
inflaton field, except that in that case the right-hand side of this
expression is $16\epsilon$, in terms of the slow-roll parameter 
$\epsilon\equiv -\dot{H}/H^2$. 

Because tensor modes would have decoupled completely from everything
else, one does not know about these fluctuations (1) whether they were
produced in the trans-Planckian region, (2) whether they emerged into the 
semi-classical universe at the same fixed length scale $\lambda_0$ as 
the scalar perturbations, or (3) whether they were even generated after 
the Planck time. If they did emerge at a fixed scale $\lambda_0^T$,
they would likely have a near-scale free spectrum, like the curvature 
perturbations, but one could not predict their power relative to 
that of the scalar fluctuations without knowing something about the 
scale at which they emerged \cite{Hollands:2002}. It may already be
possible to eliminate the first possibility, since the scaling
for $\Theta_k$ and $h_k$, through their definition in terms of
the modes $u_k$ and $v_k$, would suggest a ratio of tensor to
scalar power exceeding current upper limits. Indeed, if we adopt
the value $r\lesssim 0.11$ as the most recent observational constraint,
our model would require 
\begin{equation}
\lambda_0^T\gtrsim 12 \lambda_0\;,
\end{equation}
meaning that any gravity waves generated in this picture would
also have been produced at the GUT scale. In the case of inflation,
the slow-roll parameter $\epsilon$ is a direct probe into the
energy scale of the inflaton field, and it is generally understood
that if $r\gtrsim 0.01$, then the inflaton potential has a value
$V^{1/4}\sim (r/0.01)^{1/4}10^{16}$ GeV, which itself lies in the
GUT energy range. In some ways, this convergence of ideas is rather
promising for the quantum-perturbation model for the origin of 
fluctuations in the cosmic fluid, since it suggests that the
physics (as we know it) of scalar fields in the early Universe is 
rather tightly constrained, and perhaps the simplest extensions
to the standard model are on the horizon.

\section{Conclusions}\label{conclusions}
One of the strongest arguments in favor of the freeze-out mechanism
during inflation is the coherence of the observed CMB fluctuations
\cite{Dodelson:2003}. Curvature perturbations eventually
source density fluctuations that evolve under the influence of gravity
and pressure to produce the CMB inhomogeneities and subsequent large-scale
structure. If one reasonably supposes that recombination happens
instantaneously (at least in comparison to the evolutionary timescale),
then fluctuations with different wavelengths influence the surface of
last scattering at different phases in their oscillations. However, if
all Fourier modes of a given wavenumber have the same phase, then they
interfere coherently, resulting in a CMB spectrum with clearly
defined peaks and troughs. Without this coherence, the various modes would
all combine to produce white noise.

With inflation, all the mode phases are set when the fluctuations exit
the horizon, which therefore remain coherent upon subsequent re-entry.
Something very similar to this happens with the numen field, since all
the scalar modes of a given wavenumber $k$ emerge into the semi-classical
universe at the same scale $\lambda_0$ and, therefore, at the same
time $t_k$. These modes are super-horizon and frozen during the subsequent 
expansion. They eventually source matter and radiation fluctuations with
the same phase when the numen field decays into standard-model particles.
So the mechanism for generating coherence of the numen modes is very
similar to that of the inflaton field, though perhaps a little simpler
since it requires fewer steps and is a natural extension of the 
wavenumber-dependence of the emergence of these modes, unlike those
associated with the inflaton field, which are considered to have
been born at arbitrarily early (pre-Planckian) times \cite{Brandenberger:2002}.

The suggestion is sometimes made that Planck-era physics may eventually
be studied with the CMB. In the $R_{\rm h}=ct$ universe, this idea is
more than mere speculation. Indeed, as we have shown in this paper,
CMB fluctuation scales and amplitudes are preserved at the values
they had as they emerged out of the Planck domain. In particular,
the identification of the angular scale of the CMB inhomogeneities
with the Planck length is a strong factor in favor of this cosmology,
particularly since the Universe would have expanded by over 60 orders
of magnitude between the Planck and recombination times. One cannot
completely rule out a coincidence such as this, though the probability
of its occurrence is extremely small. So the connection between the 
CMB and Planck scales is already clearly defined in $R_{\rm h}=ct$.

We have also seen that if matter in the early universe was dominated
by a single scalar field, then its potential in this model is known
precisely (and given in Equation~18). This result may motivate
further exploration of Kaluza-Klein cosmologies, string theories, and
supergravity, in which exponential potentials such as these are well
justified. In concert with constraints imposed by CMB observations,
particularly the value of the scalar spectral index $n_s$, there is
therefore hope that new physics may emerge with relevance to the
trans-Planckian domain.

\section*{Acknowledgments}

It is a pleasure to acknowledge helpful discussions with Bob Wald, Sean
Fleming, Robert Caldwell, Daniel Sudarsky, and Robert Brandenberger. Some
of this work was carried out at Purple Mountain Observatory in Nanjing,
China, and  was partially supported by grant 2012T1J0011 from The Chinese
Academy of Sciences Visiting Professorships for Senior International Scientists.

\end{document}